\documentclass[twocolumn,reprint,prd,aps,tighten,nofootinbib,amssymb,preprintnumbers,showpacs,superscriptaddress]{revtex4-1}
\pdfoutput=1
\usepackage{graphicx}
\usepackage{epsfig,amssymb,amsmath}
\usepackage{slashed, color}
\usepackage{hyperref}
\usepackage{gensymb}
\usepackage[normalem]{ulem}
\usepackage[utf8]{inputenc}
\usepackage[export]{adjustbox}

\newcommand{\beq}{\begin{equation}}
\newcommand{\eeq}{\end{equation}}
\newcommand{\bea}{\begin{eqnarray}}
\newcommand{\eea}{\end{eqnarray}}
\newcommand{\bear}{\begin{array}}
\newcommand {\eear}{\end{array}}
\newcommand{\bef}{\begin{figure}}
\newcommand {\eef}{\end{figure}}
\newcommand{\bec}{\begin{center}}
\newcommand {\eec}{\end{center}}

\def\tri{\triangle}

\newcommand{\lsim}{
\mathrel{\hbox{\rlap{\hbox{\lower4pt\hbox{$\sim$}}}\hbox{$<$}}}}
\newcommand{\gsim}{
\mathrel{\hbox{\rlap{\hbox{\lower4pt\hbox{$\sim$}}}\hbox{$>$}}}}

\begin{document}
\draft
\tighten
\preprint{CTPU-PTC-18-18}
\preprint{KIAS-P19027}

\title{\large \bf Gamma-ray spectral modulations induced by \\
photon-ALP-dark photon oscillations }

\author{Kiwoon Choi}
\thanks{kchoi@ibs.re.kr}
\affiliation{Center for Theoretical Physics of the Universe,  Institute for Basic Science, Daejeon 34051, South Korea}

\author{Sangjun Lee}
\thanks{simflight@kaist.ac.kr}
\affiliation{Department of Physics, KAIST, Daejeon 34141, South Korea}
\affiliation{Center for Theoretical Physics of the Universe,  Institute for Basic Science, Daejeon 34051, South Korea}

\author{Hyeonseok Seong}
\thanks{sbravos@kaist.ac.kr}
\affiliation{Department of Physics, KAIST, Daejeon 34141, South Korea}

\author{Seokhoon Yun}
\thanks{SeokhoonYun@kias.re.kr}
\affiliation{Department of Physics, KAIST, Daejeon 34141, South Korea}
\affiliation{Center for Theoretical Physics of the Universe,  Institute for Basic Science, Daejeon 34051, South Korea}
\affiliation{School of Physics, Korea Institute for Advanced Study, Seoul 02455, Korea}

\begin{abstract}

Recently it has been noticed that the Fermi-LAT data of gamma-rays from some galactic pulsars and supernova remnants reveal  spectral modulations that might be explained by  the conversion of photons to ALPs (axion-like particles) induced  by the conventional ALP coupling to photon    
in the presence of galactic  magnetic fields.
 However the corresponding ALP mass and coupling are in a severe tension with the observational constraints from CAST, SN1987A, and other gamma-ray observations.  Motivated by this, 
we examine an alternative possibility that those spectral modulations
are explained by other type of ALP coupling
 involving both the ordinary photon  and a massless dark photon, 
 when nonzero background dark photon gauge fields 
are assumed. 
We  find that our scheme results in oscillations among the photon, ALP, and dark photon, which can explain  the gamma-ray spectral modulations of galactic pulsars or supernova remnants, while satisfying the known observational constraints.   

\end{abstract}

\pacs{}
\maketitle

\section{Introduction} 

Light axion-like particles (ALPs) and hidden $U(1)$ gauge bosons have been widely discussed as  well-motivated candidates for physics beyond the Standard Model of particle physics  \cite{Essig:2013lka}.
An appealing feature of those particles is that  their lightness  is protected from  unknown UV physics by symmetry.
They are also a good candidate for dark matter, and
can result in a variety of other astrophysical and cosmological consequences.
One of such phenomena associated with ALP is the photon to ALP conversions  \cite{Raffelt:1987im} in the presence of background magnetic fields, which may cause spectral modulations of X-rays \cite{Wouters:2013hua,Berg:2016ese,Conlon:2017qcw,Marsh:2017yvc} or gamma-rays \cite{Majumdar:2018sbv,Xia:2018xbt,TheFermi-LAT:2016zue,Malyshev:2018rsh,Abramowski:2013oea} from pointlike sources, spectral distortion of CMB \cite{Mirizzi:2005ng,Tashiro:2013yea,Mirizzi:2009nq,Mukherjee:2018oeb}, change of the cosmic opacity of intergalactic medium \cite{Simet:2007sa,DeAngelis:2007dqd,DeAngelis:2008sk,SanchezConde:2009wu,Avgoustidis:2010ju,Dominguez:2011xy,Horns:2012kw,Galanti:2015rda,Tiwari:2016cps,Kohri:2017ljt}, etc. 

Recently it has been noticed that the Fermi-LAT data of high energy gamma-rays from some galactic pulsars \cite{Majumdar:2018sbv} and supernova remnants \cite{Xia:2018xbt} reveal an intriguing spectral modulation (or irregularity) which might be explained by the photon to ALP conversions  caused by the coupling
\bea
\label{g_AA}
\frac{1}{4}g_{a\gamma\gamma}a F_{\mu\nu}\tilde{F}^{\mu\nu}=-g_{a\gamma\gamma}a\vec{E}\cdot \vec{B}\eea 
in the presence of galactic magnetic fields, where $F_{\mu\nu}=\partial_\mu A_\nu-\partial_\nu A_\mu$ is the gauge field strength of the ordinary $U(1)_{\rm em}$ gauge field $A_\mu$, and $\tilde F^{\mu\nu}=\frac{1}{2}\epsilon^{\mu\nu\rho\sigma}F_{\rho\sigma}$ is its dual.
However the ALP mass $m_a\simeq {\rm few}\times 10^{-9} \,{\rm eV}$ and the coupling $g_{a\gamma\gamma}\simeq {\rm few}\times 10^{-10}\, {\rm GeV}^{-1}$, which are required to explain those modulations, are in a severe tension with the CAST bound $g_{a\gamma\gamma}< 6.6\times 10^{-11} \,{\rm GeV}^{-1}$ \cite{Anastassopoulos:2017ftl}, as well as with the absence  of gamma-ray bursts  associated with  SN1987A \cite{Payez:2014xsa} and the Fermi-LAT data of gamma-rays from the galactic nucleus NGC 1275 in the center of the Perseus cluster  \cite{TheFermi-LAT:2016zue,Malyshev:2018rsh}.

Motivated by this observation, in this paper we examine an alternative scenario to explain those gamma-ray spectral modulations by means of the photon-ALP-dark photon oscillations caused by the  ALP coupling \cite{Kaneta:2016wvf,Kaneta:2017wfh}
 \bea
\label{g_AX}\frac{1}{2}g_{a\gamma\gamma^\prime}a X_{\mu\nu}\tilde F^{\mu\nu}=-g_{a\gamma\gamma^\prime}a\Big(\vec{E}\cdot \vec{B}_X+\vec{B}\cdot\vec{E}_X\Big)
\eea
in the presence of background $\langle  B_X\rangle$ and/or
$\langle  E_X\rangle$,
where $X_{\mu\nu}=\partial_\mu X_\nu-\partial_\nu X_\mu=(\vec E_X, \vec B_X)$
is the gauge field strength of a {\it massless} dark photon gauge field $X_\mu$.
Like the ordinary ALP coupling (\ref{g_AA}),  the above ALP coupling involving both the ordinary photon and dark photon generically exists once a dark photon field is introduced in the theory.
As we will see,  this coupling induces the conversion of gamma-rays to ALPs or dark photons, which can successfully explain the gamma-ray spectral modulations noticed in \cite{Majumdar:2018sbv} and \cite{Xia:2018xbt}  for certain values of the  ALP parameters and background dark photon gauge field in the range  
\bea
\label{alp_parameter}
 m_a  &\simeq & (3-7)\times 10^{-9}  \, {\rm eV},\nonumber\\
 g_{a\gamma\gamma^\prime} &\simeq & 10^{-11}-10^{-10} \,{\rm GeV}^{-1}, \\
 \langle  B_X\rangle &\simeq & 1-10\,\mu{\rm G}.\nonumber\eea 
As we will show later, contrary to the scheme based on the ordinary ALP coupling $g_{a\gamma\gamma}$, our scheme  can be compatible with all the available constraints on the scheme.

A key ingredient of our scenario is the existence of a background dark photon gauge fields $\langle B_X\rangle$ and/or $\langle E_X\rangle$ with a strength comparable to (or bigger than) the galactic magnetic fields  $\sim 1 \,\mu{\rm G}$ and  a coherent length longer than the typical galactic size $\sim 1$ kpc.  If they were generated in the early universe, such background dark photon gauge fields may result in
a dangerous distortion of CMB due to the resonant conversion of CMB photons to axions, which would occur when  the effective photon mass $m_\gamma(t)$ in the early universe crosses the ALP mass $m_a$ \cite{Mirizzi:2009nq,Tashiro:2013yea,Mukherjee:2018oeb}. To avoid this problem, we need to generate the background dark photon gauge fields at a late time
with $m_\gamma(t) < m_a$. In this paper, we will introduce an explicit model \cite{Anber:2009ua,Barnaby:2011qe,Agrawal:2017eqm,Kitajima:2017peg,Choi:2018dqr,Agrawal:2018vin,Dror:2018pdh,Co:2018lka,Bastero-Gil:2018uel}
for such a late generation of the background dark photon gauge fields, involving an additional ultra-light ALP $\phi$ with $m_\phi\lesssim 10^{-27}$ eV, whose late coherent oscillations cause a tachyonic instability of the dark photon gauge fields and amplify the vacuum fluctuations to the desired background fields $\langle B_X\rangle\sim\langle E_X\rangle \gtrsim 1 \, \mu{\rm G}$.

This paper is organized as follows. In Sec. \ref{sec:sec2} , we discuss the photon-ALP-dark photon oscillations induced by the coupling (\ref{g_AX}) in the presence of background $\langle  B_X\rangle$ and/or $\langle  E_X\rangle$. In Sec. \ref{sec:sec3},  we examine if the gamma-ray spectral modulations noticed in \cite{Majumdar:2018sbv} and \cite{Xia:2018xbt} can be explained by the conversion of gamma-rays to ALPs or dark photons, which results from 
the photon-ALP-dark photon oscillations. In Sec. \ref{sec:sec4}, we discuss the observational constraints on the ALP coupling (\ref{g_AX}) and examine if our scenario can be compatible with those constraints. Obviously the ALP coupling $g_{a\gamma\gamma^\prime}$ is free from the bound from helioscope axion search experiments such as  CAST.  Yet it should satisfy the bound from stellar evolution \cite{Raffelt:1996wa}, as well as another bound from the non-observation of gamma-ray bursts associated with SN1987A \cite{Payez:2014xsa}. 
We also need to generate the background dark photon gauge  fields at a late time with  $m_\gamma(t)<m_a$ to avoid a dangerous distortion of CMB \cite{Mirizzi:2009nq,Tashiro:2013yea,Mukherjee:2018oeb}.  In the later part of Sec \ref{sec:sec4}, we present an explicit model for such a late generation of the background dark photon gauge fields.
Sec. \ref{sec:sec5} is the conclusion.

\section{Photon-ALP-dark photon oscillations
\label{sec:sec2}}

Our scheme is based on a model involving a light ALP  and also a massless hidden $U(1)_X$ gauge boson $X_\mu$, which is dubbed the dark photon in this paper \cite{Essig:2013lka}.
We assume that there is no light $U(1)_X$-charged matter fields. 
Here we are interested in the photon-ALP-dark photon oscillations induced by the ALP coupling \cite{Kaneta:2016wvf,Kaneta:2017wfh}
\bea
\frac{1}{2}g_{a\gamma\gamma^\prime}a X_{\mu\nu}\tilde F^{\mu\nu}=-g_{a\gamma\gamma^\prime}a\Big(\vec{E}\cdot \vec{B}_X+\vec{B}\cdot\vec{E}_X\Big)\eea
in the presence of nonzero background dark photon gauge fields, $\langle  E_X\rangle$ and/or $\langle  B_X\rangle$, as well as an ordinary background  magnetic field $\langle B\rangle$.
Since there is no $U(1)_X$-charged matter, $\langle  E_X\rangle$ can be non-vanishing and then
the equations of motion of $X_\mu$  suggest  $\langle  E_X\rangle\sim\langle B_X\rangle$ \cite{Choi:2018dqr}.

Generically there can be other  ALP couplings such as
${g_{a\gamma\gamma}}aF\tilde{F}$ and ${g_{a\gamma^\prime\gamma^\prime}}aX\tilde{X}$.
For simplicity, here we will assume that $g_{a\gamma\gamma^\prime}$ dominates over
$g_{a\gamma\gamma}$ and $g_{a\gamma^\prime\gamma^\prime}$ as much as we need.  
Note that in many cases  the ALP couplings to gauge fields are quantized as they are generated by the loops of fermions  carrying quantized charges of the $U(1)_{\rm PQ}$, $U(1)_{\rm em}$ and $U(1)_X$, where $U(1)_{\rm PQ}$ denotes the (spontaneously broken) global $U(1)$ symmetry associated with the ALP field $``a"$. In those cases, one can arrange the relevant PQ and gauge charges appropriately to get the desired pattern of ALP couplings such as  $|g_{a\gamma\gamma^\prime}|\gtrsim  |g_{a\gamma^\prime\gamma^\prime}|\gg
|g_{a\gamma\gamma}|$. Alternatively, one may use the clockwork mechanism \cite{Choi:2014rja,Kaplan:2015fuy,Choi:2015fiu,Giudice:2016yja} to generate such pattern of ALP couplings 
 as was done in \cite{Higaki:2015jag,Agrawal:2017cmd,Farina:2016tgd} to get an axion-photon coupling much stronger than the axion-gluon coupling.

To examine the oscillations among the photon, dark photon and ALP, let us choose the temporal gauge $A_0=X_0=0$. Then  
the equations of motion describing the propagation of small field fluctuations of $\vec A$, $\vec X$ and $a$ in the background $\langle \vec B\rangle, \langle \vec B_X\rangle$ and $\langle \vec E_X\rangle$  are given by
\bea
\label{wave-eq}
&&\hskip -0.4cm \partial_\mu\partial^\mu \vec{A} = g_{a\gamma\gamma^\prime} \left( \langle \vec{B}_X\rangle\partial_ta  -  \langle\vec{E}_X\rangle\times \nabla a\right), \nonumber \\
&& \hskip -0.4cm \partial_\mu\partial^\mu \vec{X} = g_{a\gamma\gamma^\prime}  \langle \vec{B}\rangle\partial_t a,  \\
&& \hskip -0.4cm \partial_\mu\partial^\mu a +m_a^2 a =
 - g_{a\gamma\gamma^\prime}\left(\langle\vec{B}\rangle\cdot \partial_t\vec{X}+\langle \vec{B}_{X}\rangle\cdot \partial_t\vec{A}\right.\nonumber\\
&& \hskip 3.5cm \left. -\langle \vec{E}_X\rangle\cdot \nabla\times \vec{A}\right). \nonumber
\eea
From the above wave equations, we note that the following transverse background fields induce oscillations among the photon, dark photon and ALP:
\bea
\label{bacground_BX}
\vec B_T &=&\langle \vec B\rangle -\hat k (\hat k\cdot\langle \vec B\rangle),
\nonumber \\  
\vec B_{XT} &=&\langle \vec B_X\rangle -\hat k (\hat k\cdot\langle \vec B_X\rangle) -\hat k\times \langle \vec E_X\rangle ,
\eea
where  $\vec k$ is the momentum of the propagating particles and $\hat k=\vec k/|\vec k|$.
In the relativistic limit,
the corresponding propagation equation  is well approximated by  \cite{Raffelt:1987im}
 \bea
 \label{propagation_eq}
\Big[
w+i\partial_z - {\cal M}
\Big]
\left( \begin{tabular}{c}
$A_{\|}$\\
$X_{\|}$\\
$a$
\end{tabular}
\right) = 0 \, ,
\eea
where $\omega=|\vec k|$ and
 $z$ is the spacial coordinate along the direction $\hat k$.
Here $A_{\|}$ and $X_{\|}$ denote the polarization of $\vec A$ and $\vec X$ in the direction of
$\vec B_{XT}$ and $\vec B_T$, respectively:
\bea
A_{\|} =\frac{\vec B_{XT}\cdot \vec A}{B_{XT}}, \quad X_{\|}=\frac{\vec B_T \cdot \vec X}{B_T},\eea
and
\bea
{\cal M}=
\left( \begin{tabular}{ccc}
$\frac{\omega_{\rm pl}^2}{2\omega}$ & $0$ & $\frac{g_{a\gamma\gamma^\prime}B_{XT}}{2}$ \\
$0$ & $0$ & $\frac{g_{a\gamma\gamma^\prime}B_T}{2}$ \\
$\frac{g_{a\gamma\gamma^\prime}B_{XT}}{2}$ & $\frac{g_{a\gamma\gamma^\prime}B_T}{2}$ & $\frac{m_a^2}{2\omega}$
\end{tabular}
\right),\eea
where  $\omega_{\rm pl}^2=4\pi\alpha n_e/m_e$ is the plasma frequency in a medium with the electron number density
$n_e$.


Here we are mostly interested in the conversion of initial unpolarized photon to ALP or dark photon in the limit when the plasma frequency is negligible:
\bea
\omega_{\rm pl}^2\ll {\rm min}\left( m_a^2, \, g_{a\gamma\gamma^\prime}\omega \sqrt{B_T^2+B_{XT}^2}\right),  \eea
which turns out to be the case for our scenario to explain the gamma-ray spectral modulations noticed in \cite{Majumdar:2018sbv} and \cite{Xia:2018xbt}.
To get some insights on the photon-ALP-dark photon oscillations, let us  consider a simple situation  that 
all background fields  have a coherent length which is large enough to treat
them  as constants over the photon propagation distance $d$.
It is then straightforward to find that the conversion probabilities in such case are given by
 \bea
 \label{conversion_prob}
P_{\gamma\rightarrow a} =P_{a\rightarrow \gamma}
&=&\left(\frac{B_{XT}^2 }{B_{XT}^2+B_T^2}\right)\left(\frac{\omega^2}{\omega^2+\omega_c^2}\right) \sin^2 \frac{\tri_{\rm osc} d}{2}, \nonumber \\
P_{\gamma\rightarrow\gamma'} =P_{\gamma^\prime\rightarrow\gamma}
&=&\frac{2B_{XT}^2 B_T^2}{(B_{XT}^2+B_T^2)^2}\left(1-\cos\frac{\tri_{a}d}{2}\cos\frac{\tri_{\rm osc}d}{2} \right.\nonumber \\
&& \hskip 0.5cm \left. -\frac{\omega_c}{\sqrt{\omega^2+\omega_c^2}}\sin\frac{\tri_{a}d}{2}\sin\frac{\tri_{\rm osc}d}{2} \right. \nonumber \\
&&\hskip 0.5cm \left. -\frac{\omega^2}{2(\omega^2+\omega_c^2)}\sin^2\frac{\Delta_{\rm osc}d}{2}\right),\eea
where $d$ is the propagation distance, and
\bea
&&\omega_c= \frac{m_a^2}{2g_{a\gamma\gamma^\prime}B_{\rm eff}},\quad
 \Delta_a={g_{a\gamma\gamma^\prime}B_{\rm eff}}\frac{\omega_c}{\omega},\nonumber\\
&&\hskip 0.5cm \Delta_{\rm osc} = g_{a\gamma\gamma^\prime}B_{\rm eff}\sqrt{1+\left(\frac{\omega_c}{\omega}\right)^2}
\eea
for
\bea
B_{\rm eff}\equiv \sqrt{B_{XT}^2+B_T^2}.\eea
From the above results, we find the following photon survival probability for the photon-ALP-dark photon oscillations induced by the ALP coupling $g_{a\gamma\gamma^\prime}$:
\bea
\label{photon_survival}
\left(P_{\gamma\rightarrow\gamma}\right)_{g_{a\gamma\gamma^\prime}} 
&=&\frac{B_{XT}^4+B_T^4}{(B_{XT}^2+B_T^2)^2}  \nonumber\\
&&\hskip -1cm -\left(\frac{B_{XT}^2}{B_{XT}^2+B_T^2}\right)^2\left(\frac{\omega^2}{\omega^2+\omega_c^2}\right)\sin^2\frac{\tri_{\rm osc}d}{2}
\nonumber \\
&&  \hskip -1cm
+\,2\left(\frac{B_{XT} B_T}{B_{XT}^2+B_T^2}\right)^2 \left( \cos\frac{\tri_{a}d}{2}\cos\frac{\tri_{\rm osc}d}{2} \right.\nonumber\\
&& \hskip -1cm \left.+\frac{\omega_c}{\sqrt{\omega^2+\omega_c^2}}\sin\frac{\tri_{a}d}{2}\sin\frac{\tri_{\rm osc}d}{2}\right) \, . \qquad
\eea

For later discussion of the possibility that the ALP coupling $g_{a\gamma\gamma^\prime}$ explains the gamma ray modulations noticed in \cite{Majumdar:2018sbv} and \cite{Xia:2018xbt},
let us compare the photon depletion caused by $g_{a\gamma\gamma^\prime}$ in background  $U(1)_X$ gauge fields  with  the
photon depletion by $g_{a\gamma\gamma}$ in ordinary magnetic fields.
As is well known \cite{Raffelt:1987im},
the photon survival probability for the photon-ALP oscillations caused by $g_{a\gamma\gamma}$ in ordinary magnetic fields is given by 
\bea
 (P_{\gamma\rightarrow\gamma})_{g_{a\gamma\gamma}} = 1-\left(\frac{\omega^2}{\omega^2+\tilde\omega_c^2}\right)\sin^2\frac{\tilde \Delta_{\rm osc}d}{2}, \label{photon_survival_1}
  \eea
  where 
  \bea
  \tilde \omega_c =\frac{m_a^2}{2g_{a\gamma\gamma} B_T},\quad
  \tilde \Delta_{\rm osc} = 
  g_{a\gamma\gamma}B_T\sqrt{1+\left(\frac{\tilde\omega_c}{\omega}\right)^2}.\eea
There are two key spectral features of the above photon survival probability.
The first is the step-down of $P_{\gamma\rightarrow\gamma}$ at $\omega\sim \tilde\omega_c$, which represents  the transition from the small mixing limit $\omega\ll \tilde \omega_c$ to the large mixing limit $\omega \gg \tilde \omega_c$. The size of step-down is given by
\bea
\Delta P  &\equiv& P_{\gamma\rightarrow \gamma}(\omega\ll \tilde \omega_c) - P_{\gamma\rightarrow \gamma}(\omega\gg \tilde \omega_c) \nonumber\\
&=&  \sin^2\left(\frac{g_{a\gamma\gamma}B_T d}{2}\right), \eea
which can be significant when
 $g_{a\gamma\gamma}B_Td \gtrsim {\cal O}(1)$. 
The second feature of $(P_{\gamma\rightarrow\gamma})_{g_{a\gamma\gamma}}$
 is the oscillatory behavior  due to $\sin^2 (\tilde \Delta_{\rm osc}d/2)$, which 
  becomes appreciable right before the step-down.  
  The spectral modulations (irregularities) noticed in \cite{Majumdar:2018sbv} and \cite{Xia:2018xbt} 
 could result from the combination of these two effects on the gamma-ray intensity around $\omega={\cal O}(1)$\,GeV.

 The photon survival probability $(P_{\gamma\rightarrow\gamma})_{g_{a\gamma\gamma^\prime}}$ of (\ref{photon_survival})
 appears to be significantly  more complicated than 
 $(P_{\gamma\rightarrow\gamma})_{g_{a\gamma\gamma}}$ of (\ref{photon_survival_1}).
  Yet, as long as $B_{XT}$ is {\it not} negligible compared to $B_T$, it reveals a similar step-down behavior  at $\omega\sim \omega_c$, as well as a similar oscillatory feature. 
To see this, let us first consider the case $B_{XT}\gg B_T$.
 In such case, the dark photon is decoupled and the resulting photon-ALP oscillation yields  
\bea
&& (P_{\gamma\rightarrow\gamma})_{g_{a\gamma\gamma^\prime}}\nonumber\\
 &&\hskip 0.3cm \simeq 1-\left(\frac{\omega^2}{\omega^2+\omega_c^2}\right) \sin^2\left(\frac{g_{a\gamma\gamma^\prime}B_{XT}d}{2}\sqrt{1+\frac{\omega_c^2}{\omega^2}}\right)\qquad
  \eea
for $B_T\ll B_{XT}$.
Obviously the above $\left(P_{\gamma\rightarrow\gamma}\right)_{g_{a\gamma\gamma^\prime}}$ mimics 
 $(P_{\gamma\rightarrow\gamma})_{g_{a\gamma\gamma}}$
 with the simple 
 correspondence 
\bea g_{a\gamma\gamma^\prime}B_{XT}\,\,\rightarrow\,\, g_{a\gamma\gamma}B_T.\eea

Even when $B_{XT}\lesssim B_T$, if it is non-negligible,  
there can be a significant step-down of $(P_{\gamma\rightarrow\gamma})_{g_{a\gamma\gamma^\prime}}$ at $\omega\sim \omega_c$. 
From (\ref{photon_survival}), one finds
\bea
&&\left(P_{\gamma\rightarrow\gamma}\right)_{g_{a\gamma\gamma^\prime}}(\omega \ll \omega_c)\nonumber\\
&&\hskip 0.5cm \simeq
1-\frac{4B_{XT}^2B_T^2}{(B_{XT}^2+B_T^2)^2}\sin^2\left(\frac{g_{a\gamma\gamma^\prime}B_{\rm eff}d}{8}\frac{\omega}{\omega_c}\right)\qquad\quad
\eea
and
\bea
&&\left(P_{\gamma\rightarrow\gamma}\right)_{g_{a\gamma\gamma^\prime}}(\omega \gg \omega_c)\nonumber\\
&&\hskip 0.5cm \simeq
1-\frac{B_{XT}^4}{(B_{XT}^2+B_T^2)^2}\sin^2 \left(\frac{g_{a\gamma\gamma^\prime}B_{\rm eff}d}{2}\right)
\nonumber \\
&&\hskip 0.8cm  -\frac{2B_{XT}^2B_T^2}{(B_{XT}^2+B_T^2)^2}\left(1-\cos\left(\frac{g_{a\gamma\gamma^\prime}B_{\rm eff}d}{2}\right)\right.\nonumber\\
&&\hskip 3.6cm \times\left.\cos\left(
\frac{g_{a\gamma\gamma^\prime}B_{\rm eff}d}{2}\frac{\omega_c}{\omega}\right)\right).\quad
\eea
When
$g_{a\gamma\gamma^\prime}B_{\rm eff}d ={\cal O}(1)$, which is the case relevant for us,
this results in the following step-down of the photon survival probability:
\bea
\Delta P &=& P_{\gamma\rightarrow \gamma}(\omega\ll  \omega_c) - P_{\gamma\rightarrow \gamma}(\omega\gg \omega_c) \nonumber \\
&\simeq &
\frac{4B_{XT}^2B_T^2}{(B_{XT}^2+B_T^2)^2} \sin^2\left(\frac{g_{a\gamma\gamma^\prime}B_{\rm eff}d}{4}\right) \nonumber\\
&&+\frac{B_{XT}^4}{(B_{XT}^2+B_T^2)^2}  \sin^2\left(\frac{g_{a\gamma\gamma^\prime}B_{\rm eff}d}{2}\right).
\eea
This step-down of the photon survival probability can be sizable enough to yield an observational consequence if $B_{XT}$ is comparable to $B_T$.

The discussions above indicate that the results of \cite{Majumdar:2018sbv} and \cite{Xia:2018xbt}, which are based on the ALP coupling $g_{a\gamma\gamma}$,
can be reproduced also by $g_{a\gamma\gamma^\prime}$ for appropriate range of ALP parameters  if the background dark photon gauge fields $E_X$ and/or $B_X$ are comparable to or even larger than the galactic magnetic fields $B\sim 1\, \mu{\rm G}$. In the next section, we examine this possibility in more detail
for specific gamma-ray sources,  the galactic pulsar J2021+3651 and the supernova remnant IC443.


\section{Gamma-ray spectral modulations
\label{sec:sec3}}

In \cite{Majumdar:2018sbv} and \cite{Xia:2018xbt}, Fermi-LAT data of gamma-rays from some galactic pulsars with a distance $d\simeq 5- 10$ kpc and supernova remnants 
with $d\simeq 1-5$ kpc were examined to see if there is any irregularity in the spectrum. It was then noticed that the data indicates spectral modulations at $\omega \sim 1\,{\rm GeV}$, which might be explained by the photon-ALP oscillations induced by an ALP coupling $g_{a\gamma\gamma}\sim{\rm few}\times 10^{-10}\, {\rm GeV}^{-1}$  in the presence of galactic  magnetic fields $B\sim 1 \,\mu{\rm G}$ for ALP mass $m_a\sim{\rm few}\times 10^{-9}\, {\rm eV}$.  However these ALP parameters are in conflict with the bound from  CAST  \cite{Anastassopoulos:2017ftl}, as well as with the non-observation of gamma-ray bursts associated with SN1987A \cite{Payez:2014xsa}.
 This motivates us an alternative scenario that
those spectral modulations are explained by the photon-ALP-dark photon oscillations induced by $g_{a\gamma\gamma^\prime}$, which were discussed in the previous section.

In the previous section, the conversion probabilities for the photon-ALP-dark photon oscillations 
have been derived when the background gauge fields are assumed to be constant. In the following, we take a more elaborate approach taking into account the spacial variation of galactic magnetic fields
 along the line of sight towards the gamma-ray sources, and solve the propagation equation (\ref{propagation_eq}) numerically.
Specifically, we 
adopt the latest model for galactic profile of the magnetic field, i.e. the Jansson and Farrar model \cite{Jansson:2012pc}, which was adopted also  in \cite{Majumdar:2018sbv} and \cite{Xia:2018xbt}.
As for the background dark photon gauge field component $B_{XT}$, we simply assume that 
it is homogeneous enough\footnote{Indeed our mechanism to generate the background dark photon gauge fields, which will be presented in the later part of the next section,  yields a coherent length
$\lambda \gg $ kpc of the produced $B_{XT}$, justifying this assumption.} to be treated as constant over the galactic distance scales of ${\cal O}(1)$ kpc.



Let us now examine in more detail if our photon-ALP-dark photon oscillation scenario can explain the observed gamma-ray spectral modulations in the spectra of PSR J2021+3651 and IC443. For this, we perform a simplified $\chi^2$ analysis over the parameter space including the ALP mass $m_a$,  ALP coupling $g_{a\gamma\gamma^\prime}$, and the fitting parameters of the intrinsic spectrum. Here the simplified  analysis means that we ignore the dispersion matrix effect which would take into account of systematic errors on the reconstruction of the true energy.
Based on the Fermi data instruction on the energy dispersion\footnote{\href{https://fermi.gsfc.nasa.gov/ssc/data/analysis/documentation/Pass8_edisp_usage.html}{/analysis/documentation/Pass8\_edisp\_usage.html}}, the instrument response effects are expected to give a few percent level of change in the spectral parameters for the energy range in our case. As the main goal of this work is to propose a new theoretical idea,
more refined $\chi^2$ analysis taking into account of the full systematic uncertainties is beyond the scope of this work.
Yet our simplified  $\chi^2$ analysis is expected to be good enough to identify the parameter region which can explain the gamma-ray spectral modulations noticed in \cite{Majumdar:2018sbv} and \cite{Xia:2018xbt}.

\begin{figure}[t]
\begin{center}
\includegraphics[width=.45\textwidth]{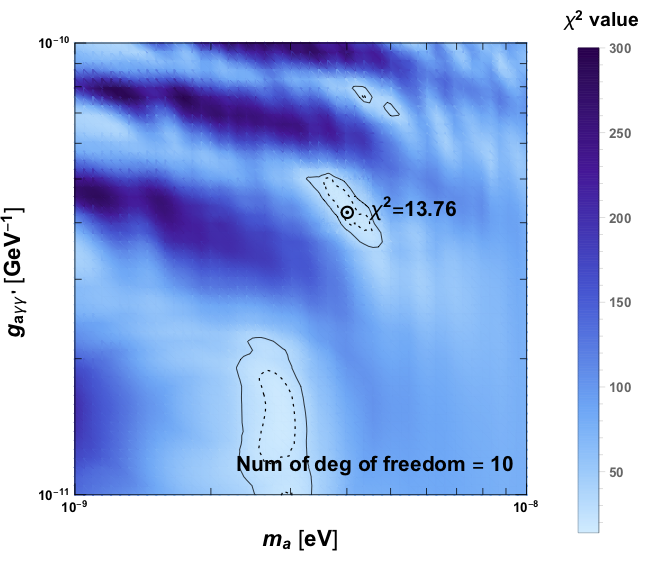}\vspace{0.3cm}
\hspace*{0.3cm}\includegraphics[width=.4\textwidth,left]{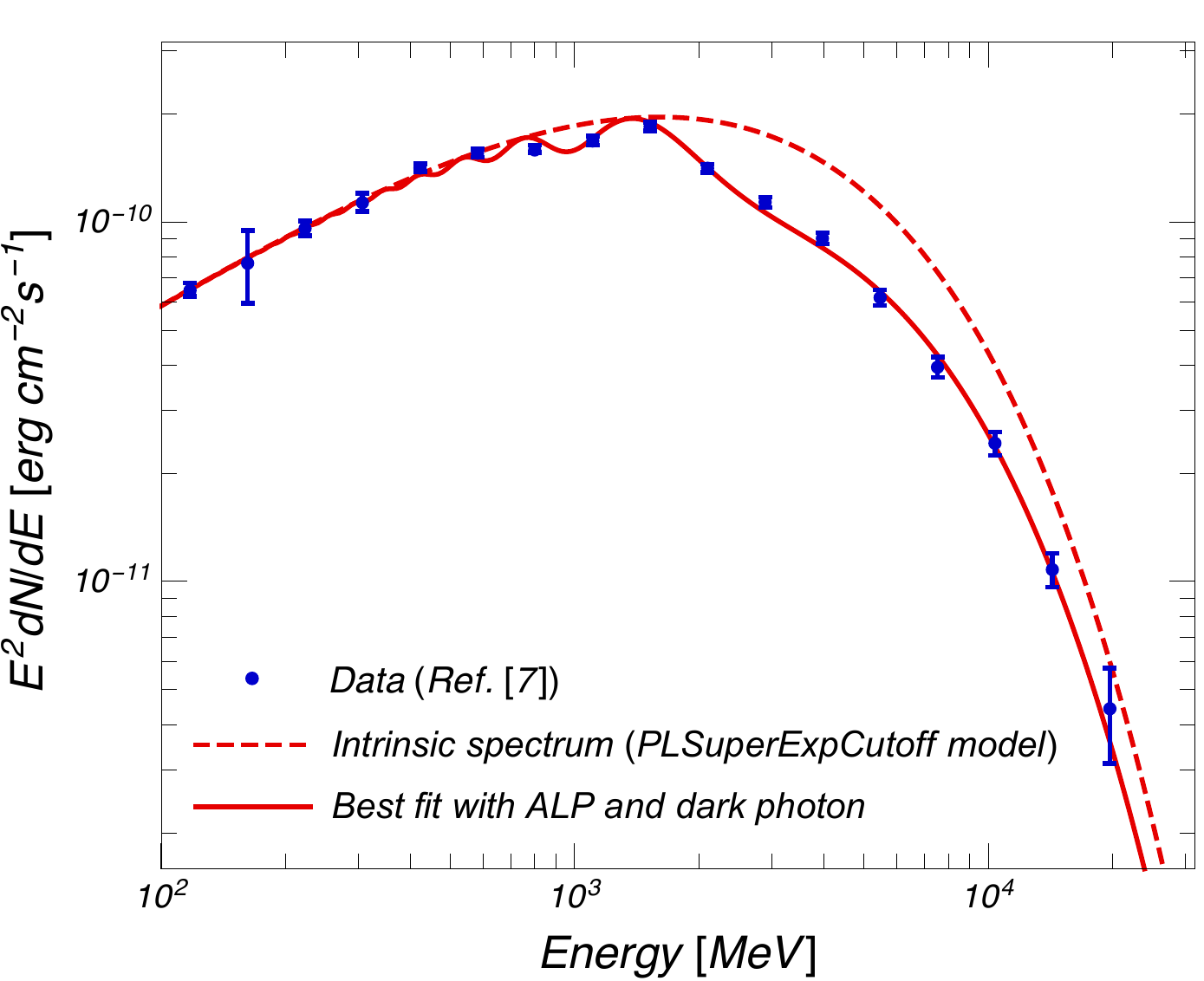}
\caption{\small {(Top) $\chi^2$ values for PSR J2021+3651 with varying $m_a$, $g_{a\gamma\gamma^\prime}$ and the intrinsic spectrum parameters. Here the background dark photon gauge field is fixed as $B_\text{XT}=6.5\mu{\rm G}$ for simplicity.  The solid (dotted) line represents the 1-$\sigma$ (2-$\sigma$) contour and $\odot$ denotes  the best fit point for the data taken from \cite{Majumdar:2018sbv}. (Bottom) Best fit spectrum
(red solid line) for $B_\text{XT}=6.5\mu{\rm G},$ $\log_{10}(m_a/\text{eV})=-8.396$, and $\log_{10}(g_{a\gamma\gamma'}/{\text{GeV}^{-1}})=-10.37$, which correspond to $\odot$ in the top panel. Blue dots with error bars denote the data taken from \cite{Majumdar:2018sbv}. For the intrinsic spectrum (red dashed line), we use
the PLSuperExpCutoff model as  \cite{Majumdar:2018sbv}. 
}}
\label{spectrum_J2021}
\end{center}
\end{figure}

Fig.~\ref{spectrum_J2021} shows the result of our  $\chi^2$ analysis for PSR J2021+3651 in our scenario. To be specific, here we assume $B_{XT}=6.5\,\mu{\rm G}$ (for the generation of such background dark photon gauge field, see sec.~\ref{sec:generation}). We  take the data from \cite{Majumdar:2018sbv} and use the same setup as \cite{Majumdar:2018sbv}  for the galactic magnetic fields (Jansson and Farrar \cite{Jansson:2012pc}) and the intrinsic spectrum  (PLSuperExpCutoff\footnote{\label{footnote:srcmodel}\href{https://fermi.gsfc.nasa.gov/ssc/data/analysis/scitools/source_models.html}{/analysis/scitools/source\_models.html}}).
We also adopt the relative systematic uncertainty of $2.4\%$ for the consistency with \cite{Majumdar:2018sbv}. Then the best fit parameters are found to be
\begin{equation}
\label{alp_parameter1}
\begin{array}{r@{}l}
 \displaystyle   m_a  &{}\simeq 4.0~\text{neV},\vspace{0.2cm}\\
 \displaystyle   g_{a\gamma\gamma'}&{}\simeq 4.3\times 10^{-11}~\text{GeV}^{-1}
\end{array}
\end{equation}
for $B_\text{XT}=6.5\mu{\rm G}$.


\begin{figure}[t]
\begin{center}
\includegraphics[width=.45\textwidth]{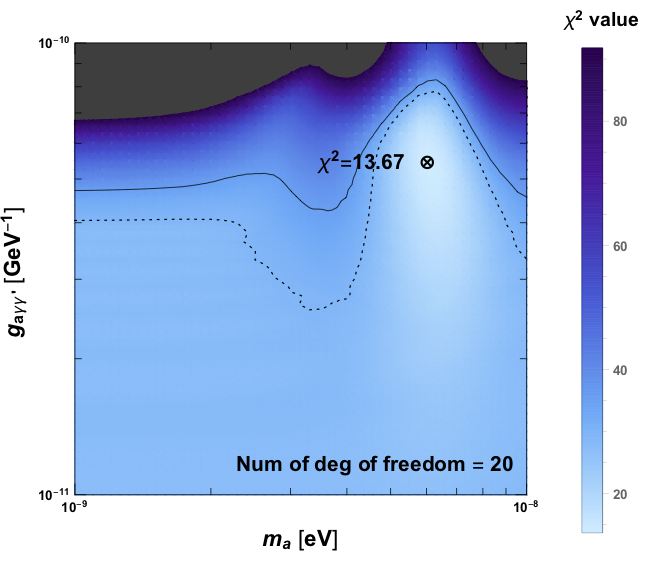}\vspace{0.3cm}
\hspace*{0.4cm}\includegraphics[width=.4\textwidth,left]{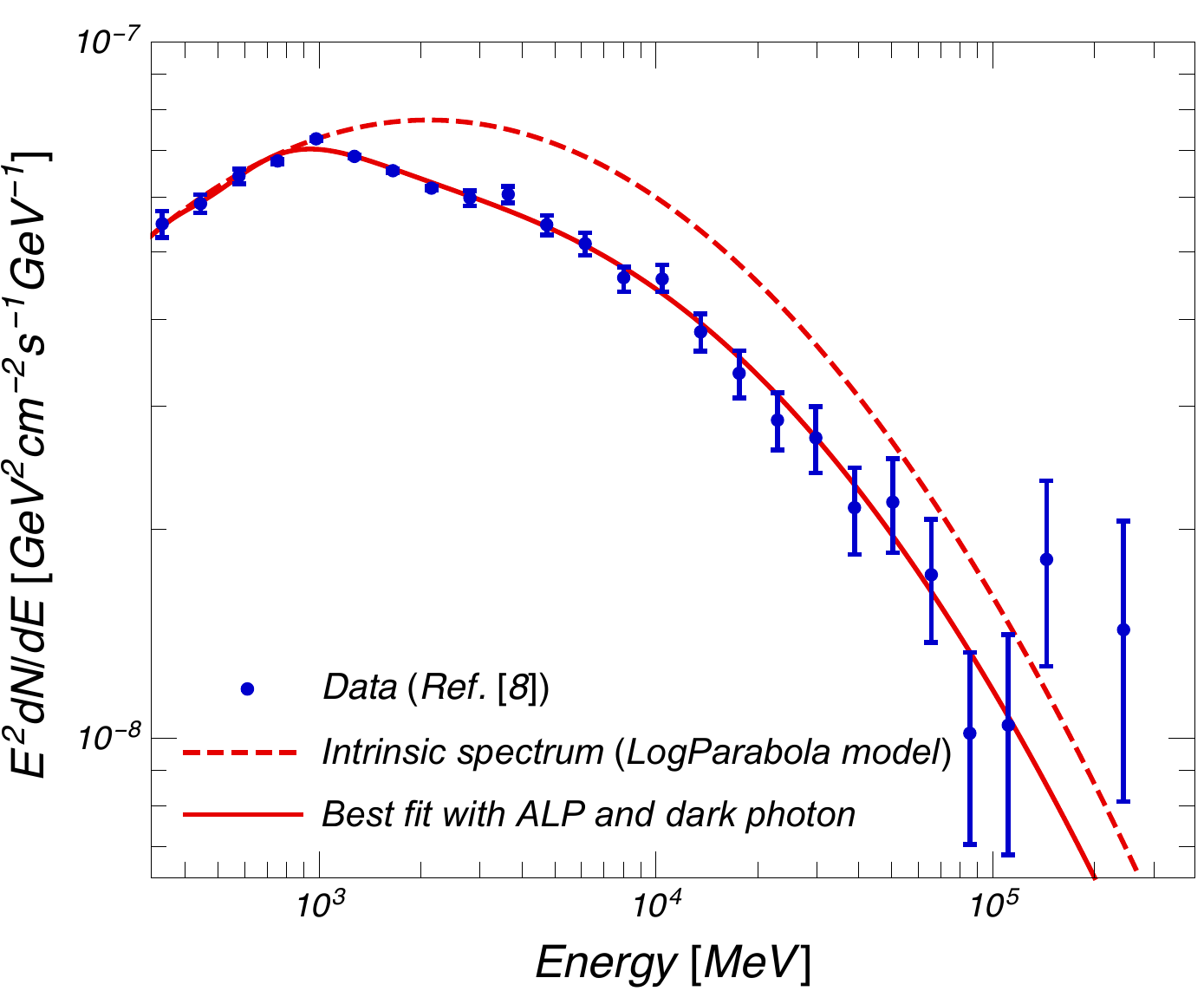}
\caption{\small {(Top) $\chi^2$ values for IC443 with varying $m_a$, $g_{a\gamma\gamma^\prime}$ and  the intrinsic spectrum parameters. Again the background dark photon gauge field is fixed as $B_\text{XT}=6.5\mu{\rm G}$,
the solid (dotted) line represents the 1-$\sigma$ (2-$\sigma$) contour,   and
$\otimes$ denotes the best fit point for the data taken from \cite{Xia:2018xbt}. Dark gray region is cut out because of large $\chi^2$ which worsens the contrast in the figure. (Bottom) Best fit spectrum (red solid line) for $B_\text{XT}=6.5\mu{\rm G},$ $\log_{10}(m_a/\text{eV})=-8.22$, and $\log_{10}(g_{a\gamma\gamma'}/{\text{GeV}^{-1}})=-10.26$, which corresponds to $\otimes$ in the top panel. Blue dots with error bars denote the data taken from \cite{Xia:2018xbt} and the LogParabola model is used for the intrinsic spectrum (red dashed line) as in \cite{Xia:2018xbt}.}}
\label{fig:SNremnant}
\end{center}
\end{figure}

In Fig.~\ref{fig:SNremnant}, we do the same $\chi^2$ analysis for IC443. We again assume $B_{XT}=6.5\mu G$ and
take the data from \cite{Xia:2018xbt}. As for galactic magnetic fields, we use  the Jansson and Farrar model \cite{Jansson:2012pc}. However the intrinsic spectrum is modelled by LogParabola function\textsuperscript{\ref{footnote:srcmodel}} and the relative systematic uncertainty is set by $3\%$ for
the consistency with \cite{Xia:2018xbt}. We then find  the best fit parameters for IC443 are given by
\begin{equation}
\label{alp_parameter2}
\begin{array}{r@{}l}
 \displaystyle   m_a &{} \simeq 6.0~\text{neV},\vspace{0.2cm}\\
 \displaystyle   g_{a\gamma\gamma'} &{} \simeq 5.5\times10^{-11}~\text{GeV}^{-1}
\end{array}
\end{equation}
for $B_\text{XT}=6.5\mu{\rm G}$.

Our results clearly show that 
the spectral modulations of galactic pulsars and supernova remnants
noticed in \cite{Majumdar:2018sbv} and \cite{Xia:2018xbt} can be explained 
by the photon-ALP-dark photon oscillations caused by the ALP coupling $g_{a\gamma\gamma^\prime}$,
with a significance similar to that of the explanation based on the ordinary ALP coupling $g_{a\gamma\gamma}$.
As will be examined in detail in the next section, unlike the explanation based on $g_{a\gamma\gamma}$, our scheme can be compatible with the known observational constraints.
It is remarkable that although the distance $d\simeq 1.5$ kpc from IC443 is  quite shorter than the distance $d\simeq 10$ kpc
 from J2021+3651, the best fit point of  ($m_a$, $g_{a\gamma\gamma^\prime}$)  to explain the spectral irregularity of IC443 is close to the best fit point of J2021+3651  for $B_{XT}=6.5$ $\mu{\rm G}$. Therefore we expect that there exists a region of
($m_a$, $g_{a\gamma\gamma^\prime}$, $B_{XT}$) which can explain the both spectral modulations with the same parameter values and large enough significance.  Finding such parameter values requires an extended $\chi^2$ analysis with varying $B_{XT}$, which is beyond the scope of this paper.
\section{Observational constraints and a late generation of the background dark photon gauge fields
\label{sec:sec4}}

In this section, we examine if our photon-ALP-dark photon oscillation scenario
discussed in the previous section 
can be compatible with the observational constraints available at present.
Note that the major problem of the previous explanation based on the ALP coupling $g_{a\gamma\gamma}$, which was suggested in
 \cite{Majumdar:2018sbv} and \cite{Xia:2018xbt}, 
 is the conflict with the bound on $g_{a\gamma\gamma}$ from CAST \cite{Anastassopoulos:2017ftl} and also  the non-observation of gamma-ray bursts from SN1987A \cite{Payez:2014xsa}\footnote{According to the recent analysis \cite{Malyshev:2018rsh}, the explanation by  $g_{a\gamma\gamma}$  is disfavored also by 
 the gamma-ray spectra of NGC 1275, although it is within the allowed region of the previous analysis \cite{TheFermi-LAT:2016zue}.}.
 
 We first note that our ALP coupling $g_{a\gamma\gamma^\prime}$ does not cause a conversion of ALP to photon in  background  magnetic field, and therefore is {\it not} constrained by helioscopic axion search experiments such as CAST.  
It is yet constrained by a variety of astrophysical and/or cosmological observations, including 
the stellar evolution, the absence of gamma-ray bursts from SN1987A, and a distortion of CMB.
In the following, we examine those constraints on $g_{a\gamma\gamma^\prime}$ to see if the 
 ALP parameters and the background dark photon gauge fields which would explain the gamma-ray modulations noticed in
 \cite{Majumdar:2018sbv} and \cite{Xia:2018xbt} can pass the observational constraints. In the later part of this section, we
introduce an explicit model to generate the background dark photon gauge field $\langle B_X\rangle$ and/or $\langle E_X\rangle$, which is a key ingredient of our scheme.

\subsection{Stellar evolutions}

Let us start with the constraint from stellar evolution. 
If light hidden sector particles are produced  and successfully emitted from the core of star, such additional energy loss mechanism can affect the properties of the stellar object such as the core mass, luminosity, and lifetime \cite{Raffelt:1996wa}.
In an ionized plasma such as the core of star, the excitations of the electromagnetic field possess a different dispersion relation due to the coherent interaction with medium, and the photon obtains an effective mass, the plasmon mass $\omega_{\rm pl}= \sqrt{4\pi \alpha n_e/m_e}$. In the presence of the ALP coupling $g_{a\gamma\gamma^\prime}$,  
 the massive plasmons  decay into ALP and dark photon
through the coupling $g_{a\gamma\gamma^\prime}$.
 The corresponding decay width is given by 
\bea
\Gamma_{\rm pl} = \frac{g_{a\gamma\gamma'}^2 }{96\pi} \frac{\omega_{\rm pl}^4}{\omega} \left[ \beta\left(1, \frac{m_a}{\omega_{\rm pl}}\right)\right]^{3/2}, 
\label{eq:PlasmonDecay}
\eea
where $\beta (x,y) = x^4 + y^4  - 2x^2y^2$.
This results in the following  energy loss rate per volume:
\bea
Q_{\rm pl} = \frac{2}{2\pi^2} \int_0^\infty dk k^2 \frac{\omega \Gamma_{\rm pl}}{e^{\omega/T_c}-1} \, ,
\label{energy_loss_rate}
\eea
where $T_c$ is the core temperature. Here only the contributions from the decays of transverse plasmon modes are considered as
the longitudinal modes have a negligible number density.

The most stringent constraint associated with the above plasmon decays comes from the observationally inferred core mass at the tip of  red giant.
Any extra cooling during the red-giant phase causes a delay of the helium ignition until the core mass at the helium flash reaches  an adequate value, i.e. an increment of the core mass \cite{Raffelt:1989xu}. On the other hand, 
in view of the brightness at the tip of the red-giant branch in globular clusters, the core mass at the helium flash cannot exceed its standard value by more than 5\% \cite{Raffelt:1996wa}.
Imposing this condition, the energy loss by plasmon decays is constrained as
\bea
 \frac{Q_{\rm pl}}{\rho} \lesssim 10 ~ {\rm erg} ~ {\rm g}^{-1} ~ {\rm s}^{-1},
\eea
which results in the following upper bound:
\bea
g_{a\gamma\gamma^\prime} \lesssim 5 \times 10^{-10} ~ {\rm GeV}^{-1} \quad \mbox{for }
m_a, m_X \lesssim 10 \, {\rm keV},\quad
\eea 
where $\rho$ denotes the energy density of core, and $m_X$ denotes the dark photon mass which is assumed to be zero in our case.

\begin{figure}[t]
\begin{center}
\includegraphics[width=.43\textwidth]{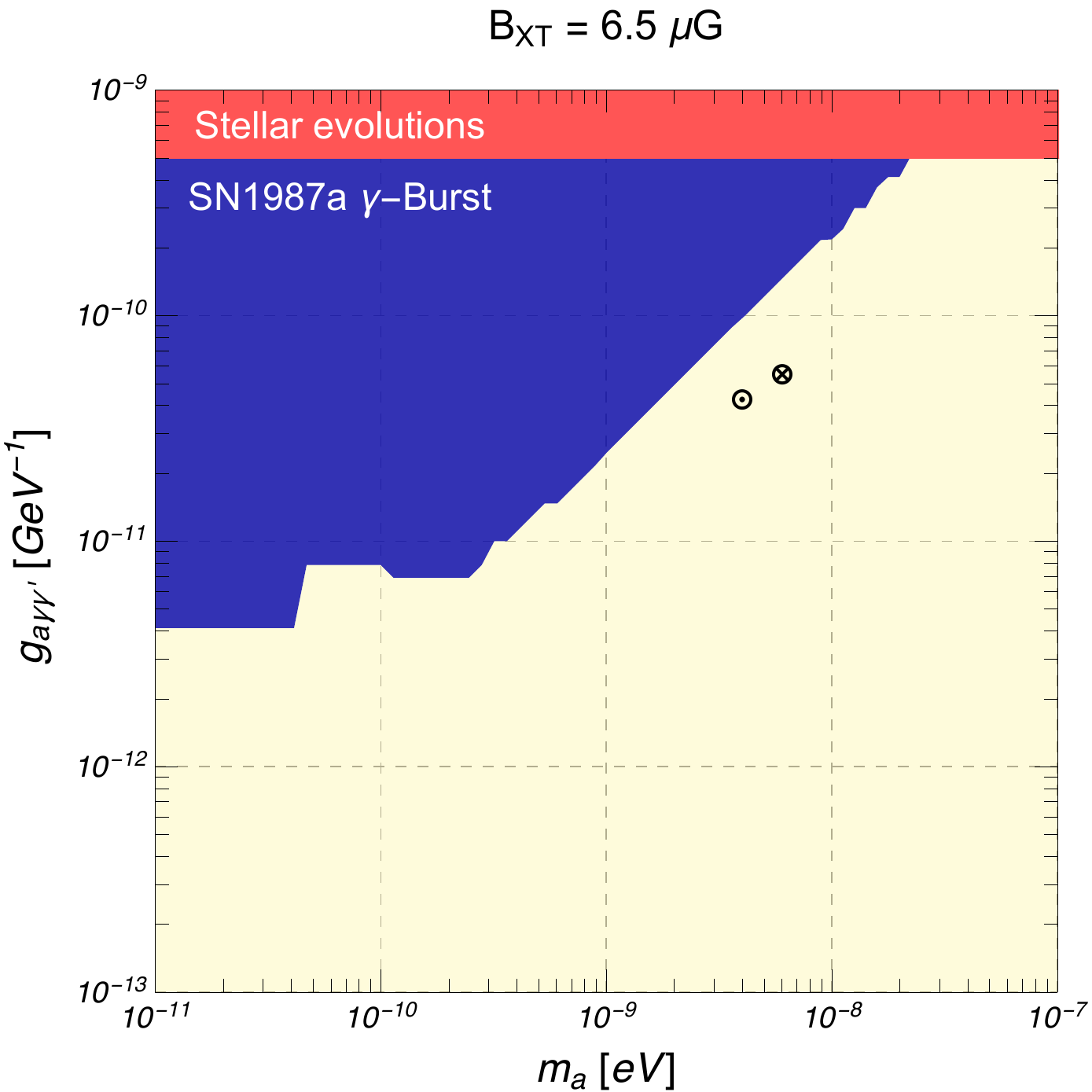}\vspace{0.3cm}
\caption{\small Red and blue colored regions in the space of $(m_a,\,g_{a\gamma\gamma^\prime})$ are excluded by stellar evolution and the absence of $\gamma$-ray bursts associated with  SN1987A for the background dark photon gauge field $B_{XT} = 6.5\,\mu {\rm G}$. 
The best fit values  of $m_a$ and $g_{a\gamma\gamma^\prime}$ for the galactic pulsar PSR J2021+3651 and the supernova remnant IC443 given in (\ref{alp_parameter1}) and (\ref{alp_parameter2}) are marked with $\odot$ and $\otimes$, respectively.}
\label{fig:GAXbound}
\end{center}
\end{figure}

\subsection{Gamma ray bursts from SN1987A}

In the presence of nonzero background dark photon fields, there can be additional constraints on $g_{a\gamma\gamma'}$.
One such example is the constraint from the  gamma-ray bursts associated with SN1987A \cite{Payez:2014xsa}.
When the neutrino signals from SN1987A were observed, the Gamma-Ray Spectrometer  on the Solar Maximum  Mission  satellite was operating to detect the $\gamma$-ray signals in the three energy bins, $4.1-6.4$ MeV, $10-25$ MeV, and $25-100$ MeV, and found 
no appreciable gamma-ray excess \cite{Chupp:1989kx}. This results in the following 95\% confident upper bounds on the gamma ray fluence
${\cal F}_{\gamma}$ over the SN neutrino duration time $t_{\rm dur}\simeq 10$ sec:
\bea
\label{fluence}
&&{\cal F}_\gamma  <  \left( 0.9,\, 0.4, \, 0.6\right) \,  \gamma\, {\rm cm}^{-2}
\eea
for $(4.1-6.4,\, 10-25, \,25-100)\,$MeV.

In our case, the ALPs and dark photons are produced by the plasmon decays in SN1987A 
whose rate is given by \eqref{eq:PlasmonDecay}.
Accepting the core-collapse supernovae model with the 18 $M_{\bigodot}$ progenitor star \cite{Fischer:2009af}, 
the plasma frequency in the degenerate core of SN1987A is given by $\omega_{\rm pl}^2 \simeq 4\alpha \mu_e^2/3\pi$, where  $\mu_e = 5.15 \, {\rm keV} (Y_e \rho_c)^{1/3}$ is the electron chemical potential for
the electron number per baryon $Y_e \simeq 0.2$ and the core density $\rho_c \simeq 3\times 10^{14}~({\rm \text{in the unit of } g \, cm}^{-3})$. The average core temperature is $T_c\simeq 30~{\rm MeV}$ which is approximately constant within the core radius $r_c \simeq 10\, {\rm km}$ during the neutrino burst duration $t_{\rm dur} \simeq 10\, {\rm sec}$.
The resulting ALP and dark photon flux per unit energy over $t_{\rm dur}$ is given by
\bea
\frac{d\Phi_{a,\gamma'}}{d \omega} = \frac{1}{4\pi d^2}\left(\frac{2\omega}{\pi}\frac{\sqrt{4\omega^2-\omega_{\rm pl}^2}}{e^{2\omega/T_c}-1}\Gamma_{\rm pl}\right)\frac{4\pi}{3}r_c^3 t_{\rm dur} \, ,\quad
\eea 
where $d \simeq 50\, {\rm kpc}$ is the distance to SN1987A.
For the ALP parameter range of our interest,  the produced ALPs and dark photons escape freely from the SN core  as their mean free path is considerably larger than $r_c$.
We then find the following total ALP and dark photon fluxes in each energy bin: \bea
&\Phi_{a,\gamma'}^{4.1-6.4\,{\rm MeV}} &\simeq 6.2 \, {\rm cm}^{-2}\times (g_{a\gamma\gamma'}/10^{-10}\,{\rm GeV}^{-1})^2, \nonumber \\
&\Phi_{a,\gamma'}^{10-25\,{\rm MeV}} &\simeq  335 \, {\rm cm}^{-2} \times (g_{a\gamma\gamma'}/10^{-10}\,{\rm GeV}^{-1})^2,\quad
 \\
& \Phi_{a,\gamma'}^{25-100\,{\rm MeV}}&\simeq  361 \, {\rm cm}^{-2} \times (g_{a\gamma\gamma'}/10^{-10}\,{\rm GeV}^{-1})^2.\nonumber\eea

Part of the ALPs and dark photons emitted from SN1987A can be converted into photons via the oscillations fueled by the background $B_{XT}$ and $B_T$, which would result in a gamma-ray burst whose differential flux is given by
\bea
\label{photon_sn}
\frac{d\Phi_{\gamma}}{d \omega} &=&  P_{a\rightarrow\gamma}\frac{d\Phi_{a}}{d \omega} + P_{\gamma'\rightarrow\gamma}\frac{d\Phi_{\gamma'}}{d \omega} \nonumber\\
 &=&  \left(P_{a\rightarrow\gamma}+P_{\gamma'\rightarrow\gamma}\right)\frac{d\Phi_{a,\gamma'}}{d \omega} \, .
\eea
For the photon-ALP oscillation caused by
 $g_{a\gamma\gamma}$, the galactic magnetic fields ($\sim \mu{\rm G}$) are the dominant source of the conversion because the intergalactic magnetic fields ($< {\rm nG}$) are negligibly small.   On the other hand, in our case 
 of the photon-ALP-dark photon oscillations induced by $g_{a\gamma\gamma^\prime}$
 for the ALP mass  $m_a={\cal O}(10^{-9}-10^{-10})$ eV, the intergalactic space between SN1987A and our Milky way galaxy where $B_{XT}\gg B_T$ mainly contributes to the conversions. Note that in our case
$B_{XT}$ is generated by a cosmological mechanism and therefore its value in intergalactic space is similar to the value inside the Milky way galaxy.  
 For such intergalactic conversions, the dark photon is decoupled and the
conversion probability 
is given by
\bea
P_{a\rightarrow\gamma}+P_{\gamma'\rightarrow\gamma}\, &\simeq&\,  P_{a\rightarrow\gamma} \nonumber\\
&\simeq&\, \frac{\omega^2}{\omega^2+\omega_c^2}\sin^2 \left(\frac{g_{a\gamma\gamma^\prime}B_{XT}\tilde d}{2}\right),\quad\eea  
where  we assumed that the coherent length of the background $B_{XT}$ is longer than
the  distance  $\tilde d\simeq $ 50 kpc
between SN1987A and the earth. Indeed, if $B_{XT}$ is generated by an additional ultra-light ALP $\phi$ at a late time with $m_\gamma(t) < m_a ={\cal O}(10^{-9}-10^{-10})$ eV, which will be presented at the end of this section, the resulting coherent length is much longer than $\tilde d \simeq 50$ kpc. 

Applying the upper bounds (\ref{fluence})  to the photon flux obtained from (\ref{photon_sn}), we can get an upper bound on $g_{a\gamma\gamma^\prime}$ for given values of $m_a$ and $B_{XT}$.  
We depict the results in Fig.~\ref{fig:GAXbound} for the background dark photon gauge field $B_{XT}=6.5 \, \mu{\rm G}$, which was  used in the previous section to explain the spectral irregularities  of the galactic pulsar PSR J2021+3651 and the supernova remnant IC443. 
In Fig.~\ref{fig:GAXbound}, the red and blue colored regions are excluded by the stellar evolution constraint and the absence of gamma-ray bursts associated with SN1987A, respectively, and $\odot$ and $\otimes$ denote
the ALP parameters which were chosen in the previous section as the benchmark  points  for PSR J2021+3651 and IC443, respectively.



\subsection{Other possible constraints}

In the previous subsections,  we discussed the constraints on $g_{a\gamma\gamma^\prime}$ from stellar evolution and the absence of $\gamma$-ray bursts associated with SN1987A.  If the background dark photon gauge fields were generated early enough, e.g.  in the early  universe with an effective photon mass $m_\gamma(t)> m_a$,  our ALP coupling  $g_{a\gamma\gamma^\prime}$ is constrained also by the absence of a significant distortion of CMB. 

It has been noticed that in the presence of   the primordial cosmological background magnetic field $B_0$, the ALP coupling $g_{a\gamma\gamma}$ can induce 
a resonant conversion of CMB photons to ALP when  $m_\gamma(t)=m_a$, which would result in a distortion of the CMB spectrum, thereby providing a bound on the combination $g_{a\gamma\gamma}B_0$ \cite{Mirizzi:2009nq,Tashiro:2013yea,Mukherjee:2018oeb}. 
The  background dark photon field $B_{XT}\gtrsim 1\,\mu{\rm G}$  in our galaxy, which is required to explain the spectral irregularities  of the galactic pulsars and supernova remnants noticed  in \cite{Majumdar:2018sbv} and \cite{Xia:2018xbt}, can not be amplified  by the galactic dynamo mechanism \cite{Davis:1999bt} and therefore should have a cosmological origin. This implies that 
a similar size of background dark photon gauge fields exist over the entire universe. 
If such cosmological  $B_{XT}$ were generated  
in the early universe when $m_\gamma(t) > m_a ={\cal O}(10^{-9})\, {\rm eV}$, so that
CMB experiences a resonant conversion to ALP  when $m_\gamma(t)=m_a$,
the results of \cite{Mirizzi:2009nq,Tashiro:2013yea,Mukherjee:2018oeb}
can be straightforwardly applied 
 to our case, yielding a bound
$g_{a\gamma\gamma^\prime} \langle B_{XT}\rangle \, <\, 10^{-14} \, {\rm GeV}^{-1}\mu {\rm G}$ for an ALP mass  $m_a={\cal O}(10^{-9})$ eV,
where 
$\langle B_{XT}\rangle$ denotes the cosmic average of $B_{XT}$.
On the other hand, as was noticed in the previous section, we need $g_{a\gamma\gamma^\prime}  B_{XT}={\cal O}(10^{-10}) \, {\rm GeV}^{-1}\mu {\rm G}$ to explain the spectral irregularities  of the galactic pulsars and supernova remnants, where now $B_{XT}$ corresponds to the local dark photon field strength in our galaxy. As the local dark photon field strength is expected to be similar to its cosmic average,  this implies that the background dark photon gauge fields should be generated
at a late time when $m_\gamma (t) <m_a ={\cal O}(10^{-9})\, {\rm eV}$, i.e. at the red-shift
\bea
\label{cmb-distortion}
z_X  \,<\, z_{\rm res} \,\simeq\,  {\rm few}\times 10^3\eea
where $z_X$ denotes the red-shift factor at the time when the background dark photon gauge fields are produced, and $z_{\rm res}$ is the red-shift factor when
$m_\gamma(t) = m_a  ={\cal O}(10^{-9})\, {\rm eV}$.

Even in such case avoiding dangerous resonant conversion, CMB distortion can occur through non-resonant conversion. The corresponding conversion probability can be easily estimated as
 \bea
 P_{\gamma\rightarrow a}^{\rm non-res}\sim \frac{g_{a\gamma\gamma^\prime}^2B_{XT}^2\omega_{\rm CMB}^2}{m_a^4}\lesssim  10^{-8},\eea
  where $\omega_{\rm CMB}$ denotes the CMB frequency and we used the best fit values of $(m_a,g_{a\gamma\gamma^\prime}, B_{XT})$ and  $z_X\lesssim 10^3$
  for the numerical estimate of the conversion probability.  The above value of $P_{\gamma\rightarrow a}^{\rm non-res}$ is well  below the COBE-FIRAS sensitivity $\sim 10^{-4}$ \cite{Fixsen:1996nj}, and therefore 
 our scenario is safe from the spectral distortion of CMB.

As the background dark photon gauge fields were produced  over the entire universe, they constitute a dark radiation whose energy density can be parametrized  in the unit of the energy density of an additional relativistic neutrino species as follows:
 \bea
\Delta N_{\rm eff}(X) \simeq 0.42 \,\frac{\langle E_X^2\rangle+ \langle B_X^2\rangle}{(1\, \mu{\rm G})^2},\eea 
 where $\langle E_X\rangle $ and $\langle B_X\rangle$ are the spatially averaged dark photon electric and magnetic fields  today, which are expected to be comparable to the local field strengths in our galaxy.
 If the dark photon gauge fields were produced before the matter-radiation equality, i.e. $z_X > z_{\rm eq}=3400$, their cosmological energy density would be constrained by the CMB power spectrum as
$\Delta N_{\rm eff}(X) \lesssim 0.3$ \cite{Aghanim:2018eyx}, suggesting the local $B_{XT} \lesssim {\cal O}(1)\,\mu{\rm G}$ in our galaxy. On the other hand, if $B_{XT}$ were produced after the matter-radiation equality (or the recombination), which might be required  to avoid a significant distortion of CMB (see (\ref{cmb-distortion})),  this bound  can be relaxed and $B_{XT}$ significantly bigger than 1 $\mu$G is allowed.
As we will see in the next subsection, if generated by an ultra-light ALP well after the matter-radiation equality, the produced $B_{XT}$ can be even as large as $\sim10\, \mu{\rm G}$ without any conflict with the known observational constraints.


The photon-ALP-dark photon oscillations induced by $g_{a\gamma\gamma^\prime}$ can result in 
spectral irregularities of other gamma-ray sources, and  non-observation of such irregularity may provide an upper bound on $g_{a\gamma\gamma^\prime}$.  Indeed it has been noticed recently that the Fermi-LAT and MAGIC data of  gamma-rays from  NGC1275 in the center of the Perseus cluster 
suggest\footnote{A previous study \cite{TheFermi-LAT:2016zue} indicates that $g_{a\gamma\gamma}={\cal O}(10^{-10}) \, {\rm GeV}^{-1}$ for $m_a={\cal O}(10^{-9})$ eV is allowed by the same Fermi-LAT data on the gamma-rays from NGC 1275, and therefore the result of \cite{Malyshev:2018rsh} should be taken with some caution.}
  $g_{a\gamma\gamma}\lesssim  10^{-12}\, {\rm GeV}^{-1}$ for an ALP mass $m_a={\cal O}(10^{-10}-10^{-9})$ eV \cite{Malyshev:2018rsh}.
However such consideration cannot be applied to our case because
 the possible spectral irregularity caused by $g_{a\gamma\gamma^\prime}$ severely depends on the detailed
profile (including the directions) of the background dark photon gauge fields\footnote{It  depends also on the profile of ordinary magnetic fields inside the Perseus cluster.} along the line of sight between the earth and NGC 1275, on which we don’t have any information.  

\subsection{Generation of the background dark photon gauge fields\label{sec:generation}}

A key ingredient of our scenario is the  background dark photon gauge fields $E_X$ and/or $B_X$ which are either comparable to or even stronger than the galactic magnetic fields $B\sim 1\,\mu{\rm G}$.  As there is no reason that the dark photon gauge fields are particularly strong in our galaxy, it is expected that a comparable size of dark photon gauge fields exist over the entire universe. An attractive mechanism to generate such  cosmological background gauge fields is to amplify the vacuum fluctuations through the tachyonic instability caused by an evolving axion-like field \cite{Anber:2009ua,Barnaby:2011qe,Agrawal:2017eqm,Kitajima:2017peg,Choi:2018dqr,Agrawal:2018vin,Dror:2018pdh,Co:2018lka,Bastero-Gil:2018uel}.  As was noticed in the previous subsection, to avoid a resonant conversion of CMB to ALP in the early universe \cite{Mirizzi:2009nq,Tashiro:2013yea,Mukherjee:2018oeb}, which would result in a dangerous distortion of CMB spectrum, those dark photon gauge fields should be generated at a late time with the corresponding  red-shift factor $z_X < z_\text{res}\simeq {\rm few}\times 10^3$ where $m_\gamma (z_\text{res}) = m_a={\cal O}(10^{-9})$ eV. If one wishes to have $E_X$ and/or $B_X$ significantly stronger than 1 $\mu{\rm G}$,  then the background dark photon gauge fields need to be generated well after the matter-radiation equality, e.g. at
$z_X < 1000$, to avoid the bound on the dark photon energy density from the CMB power spectrum \cite{Aghanim:2018eyx}.
In the following, we present an explicit model for such a late generation of the cosmological background dark photon gauge fields, involving an additional ultra-light ALP $\phi$ whose low energy effective Lagrangian is given by
\bea
{\cal L}_\phi=\frac{1}{2}\partial^\mu\phi\partial_\mu\phi-\frac{1}{2}m_\phi^2 \phi^2 -\frac{1}{4}g_{\phi\gamma^\prime\gamma^\prime} \phi X_{\mu\nu}\tilde X^{\mu\nu}.\eea

In the early universe with the Hubble expansion rate $H \gg m_\phi$, $\phi$ is frozen at its initial value $\phi_i$. Around the time when $H$ becomes comparable to $m_\phi$, more specifically at $t=t_{\rm osc}$ when $3H(t_{\rm osc})\simeq m_\phi$, the ultra-light ALP $\phi$ begins to oscillate, and evolves for a while as follows 
\bea
\phi(t) \simeq  \Big(R(t)/R(t_{\rm osc})\Big)^{-3/2}\phi_i\cos\Big(m_\phi(t-t_{\rm osc})\Big),\,\,
\eea
where $R(t)$ is the scale factor in the spacetime metric of the expanding universe:
\bea
ds^2 =dt^2-R^2(t) d\vec x^2=R^2(\tau)(d\tau^2- d\vec x^2).\eea
Then the equation of motion for the dark photon field in the momentum
space is given by
\bea
X_{k\pm}''+k(k\mp g_{\phi\gamma^\prime\gamma^\prime} \phi')X_{k\pm}= 0,
\eea
where the prime represents the derivative with respect to the conformal time $\tau$, and the subscripts $k$ and  $\pm$ denote the comoving momentum and the helicity, respectively.
 This shows
that in the oscillating background $\phi$, one of the helicity states of $X_k$ experiences a tachyonic instability for certain range
of $k$. The vacuum  fluctuations of $X_k$ in this range of
$k$ are exponentially amplified to be a stochastic classical field  \cite{Anber:2009ua,Barnaby:2011qe,Agrawal:2017eqm,Kitajima:2017peg,Choi:2018dqr,Agrawal:2018vin,Dror:2018pdh,Co:2018lka,Bastero-Gil:2018uel}.

In order for this amplification mechanism to be efficient enough in the expanding universe, the ALP coupling $g_{\phi\gamma^\prime\gamma^\prime}$ times the ALP initial value $\phi_i$ needs 
 to be large enough, for instance  
$ g_{XX}\equiv g_{\phi\gamma^\prime\gamma^\prime}\phi_i \gtrsim 70$
to generate $E_X$ and/or $B_X\gtrsim 1 \,\mu{\rm G}$ at a late time with $z_X< 1000$.  
If the amplification mechanism is efficient enough, the produced  dark photon gauge fields can affect significantly the evolution of the ALP field $\phi$, which should be taken into account to  compute the field  strength of the produced dark photon gauge field   and also the residual relic energy density of
$\phi$ \cite{Agrawal:2017eqm,Kitajima:2017peg,Choi:2018dqr}.   As was demonstrated in the recent works \cite{Kitajima:2017peg,Choi:2018dqr}, this can be achieved by a lattice calculation of the
cosmological evolution of ALP and $U(1)$ gauge field in the expanding universe.
In order to see explicitly that our mechanism can yield the desired background dark photon gauge fields,   we performed
(following \cite{Kitajima:2017peg,Choi:2018dqr,Agrawal:2018vin}) such a lattice calculation 
of the present values of the spacially averaged  dark photon gauge field strength and the relic mass density of $\phi$ for the  
parameter region
\bea  m_\phi \lesssim 10^{-27}\, {\rm eV},\,\, 100\lesssim g_{XX}\equiv g_{\phi\gamma^\prime\gamma^\prime}\phi_i \lesssim 200.\eea
We then find 
\bea
\label{exbx}
&&\langle B_X\rangle \,\Big( \simeq \langle E_X\rangle\Big) \nonumber\\
&&\hskip 0.5cm \simeq (1.2-2)\times  
\left(\frac{m_\phi}{10^{-29} \,{\rm eV}}\right)^{-1/3} \left(\frac{\phi_i}{10^{17}\,{\rm GeV}}\right) \mu{\rm G},\,\,\quad\nonumber \\
&&\Omega_\phi h^2 \simeq  10^{-4}\left(\frac{\phi_i}{10^{17}\,{\rm GeV}}\right)^2,
\eea
where the smaller (larger) field strength  is obtained for $g_{XX}=100$ (200). We find also that
the red-shift factor $z_X$ at the time ($t_X$) when the dark photon gauge fields are produced and the present coherent length $\lambda_X$ of the produced dark photon gauge fields are determined by $m_\phi$ as
\bea
\label{lxzx}
\lambda_X &\simeq& 60 \left(\frac{m_\phi}{10^{-29}\,{\rm eV}}\right)^{-1/3} \,{\rm Mpc}, \nonumber\\
1+ z_X &=&\frac{R(t_0)}{R(t_X)}\,\simeq\, 50 \left(\frac{m_\phi}{10^{-29}\,{\rm eV}}\right)^{2/3}.\eea

To explain the gamma-ray spectral modulations 
noticed in \cite{Majumdar:2018sbv} and \cite{Xia:2018xbt} through the photon-ALP-dark photon oscillations, we need a background dark photon gauge field $B_{XT}\gtrsim 1 \,\mu{\rm G}$ in our galaxy, where
$
\vec B_{XT} =\langle \vec B_X\rangle -\hat k (\hat k\cdot\langle \vec B_X\rangle) -\hat k\times \langle \vec E_X\rangle 
$.
 As noticed in the previous subsection, those dark photon gauge fields should be produced at a late time with the red-shift $z_X < 10^3$ to avoid a dangerous distortion of CMB \cite{Mirizzi:2009nq,Tashiro:2013yea,Mukherjee:2018oeb}, as well as the constraint on dark radiation from the CMB power spectrum \cite{Aghanim:2018eyx}. Our result (\ref{lxzx}) shows that such a late production is indeed realized for the ALP mass
$m_\phi < 10^{-27} {\rm eV}$. 
 
Looking into the details of the scenario, there could be additional constraints on our scheme, particularly on the relic energy densities of $\phi$ and $X$.
Possible constraints on  the relic abundance of $\phi$ were summarized in the recent report \cite{Grin:2019mub}.   According to \cite{Grin:2019mub}, for the ALP mass range $10^{-33}{\rm eV}\ll m_\phi < 10^{-27}{\rm eV}$ which is relevant for us, the most stringent constraint on the energy density $\rho_\phi$  comes from the analysis  of CMB power spectra including the effect of CMB lensing \cite{Hlozek:2017zzf}, which leads to the bound on the relic energy density of $\phi$ today:
\bea
\label{bound1}
(\Omega_\phi h^2)_{g_{XX}=0} \lesssim 2.7\times 10^{-3}\eea 
for $m_\phi={\cal O}(10^{-30}-10^{-27}){\rm ~eV}$.
Note that this has been derived for the case that there is no conversion of $\rho_\phi$ to $\rho_X$, and therefore
corresponds to the bound for the case $g_{XX}=0$ in our notation.

Yet the above bound can be used to see if our scenario with $100\lesssim g_{XX}\lesssim 200$
can be compatible with the CMB data. 
For this, let us briefly examine the evolution of $\rho_\phi$ and $\rho_X$ in our scenario.
Before the exponential amplification of $X$, but after the oscillation of $\phi$, i.e.
during the period
with \bea
R(t_{\rm osc}) < R(t) < R(t_X),\eea    $\rho_X$ is negligibly small and therefore \bea
\Big(\rho_\phi(t)+\rho_X(t)\Big)_{100\lesssim g_{XX}\lesssim 200}
\simeq \rho_\phi(t)_{g_{XX}=0}.\eea
Over a certain period right after the amplification of $X$, e.g.
for \bea
R(t_X) < R(t) < 3R(t_X),\eea both $\rho_\phi$ and $\rho_X$ show a transient behaviour due to the violent amplification of
$X$ and the strong back reaction from the amplified $X$. Yet their total energy density is bounded as
\bea
\Big(\rho_\phi(t)+\rho_X(t)\Big)_{100\lesssim g_{XX}\lesssim 200}
\lesssim \rho_\phi(t)_{g_{XX}=0}\eea
since $\rho_X\propto R^{-4}$ is more rapidly red-shifted than $\rho_\phi\propto R^{-3}$. 
In the final stage with
\bea
R(t)\gtrsim 3 R(t_X),\eea
our lattice calculation indicates that  \bea
\label{late_behavior}
\Big(\rho_\phi(t)+\rho_X(t)\Big)_{100\lesssim g_{XX}\lesssim 200}\nonumber \\
\simeq \rho_\phi(t)_{g_{XX}=0}\left(c_1+
 c_2\frac{R(t_X)}{R(t)}\right),
\eea 
where $c_1\simeq 0.3$ and $c_2\simeq 1-2$. In summary, $(\rho_\phi(t)+\rho_X(t))_{100\lesssim g_{XX}\lesssim 200}$ 
is always bounded by $\rho_\phi(t)_{g_{XX}=0}$ and its evolution  is similar to that of $\rho_\phi(t)_{g_{XX}=0}$
except for a short transient period near $t_X$. From this, barring an uncertainty of factor few, we can draw a conclusion that
our scheme for model parameters satisfying the bound (\ref{bound1}) is compatible with the CMB data.
Using (\ref{late_behavior}), we can translate (\ref{bound1}) to the bound on $\rho_\phi$ today for the case of
$100\lesssim g_{XX}\lesssim 200$, which results in the bound
 \bea
 \label{bound2}
(\Omega_\phi h^2)_{100\lesssim g_{XX}\lesssim 200}\,\lesssim 2.7 c_1\times 10^{-3} \simeq 8\times 10^{-4}\eea
for $m_\phi={\cal O}(10^{-30}-10^{-27}){\rm ~eV}$.


We can now combine our results (\ref{exbx}) and (\ref{lxzx})  with the CMB bound (\ref{bound2}). We then find that our mechanism can successfully produce the desired 
  $B_{XT}\simeq 1-10\, \mu{\rm G}$, while satisfying the known observational constraints,  
 for the ALP mass $m_\phi={\cal O}(10^{-30}-10^{-29}) \, {\rm eV}$ and the initial ALP misalignment $\phi_i={\cal O}(10^{17})$ GeV.



\section{Conclusion
\label{sec:sec5}}

In this paper, we proposed a scenario  that the spectral irregularities of gamma-rays from some galactic pulsars and supernova remnants, which were recently noticed in \cite{Majumdar:2018sbv} and \cite{Xia:2018xbt}, are explained by means of an axion-like particle (ALP) with  a mass $m_a={\cal O}(10^{-9})$ eV and the coupling $g_{a\gamma\gamma^\prime}a X_{\mu\nu}\tilde F^{\mu\nu}$ in the range $g_{a\gamma\gamma^\prime}={\cal O}(10^{-11}-10^{-10})\, {\rm GeV}^{-1}$, where $F_{\mu\nu}$ is the  ordinary electromagnetic field and $X_{\mu\nu}$ is the field strength of a massless dark photon.
A key ingredient of our scenario is the presence of cosmological background dark photon
gauge fields, $E_X\sim B_X\gtrsim 1\,\mu{\rm G}$, which can be successfully generated by an additional ultra-light ALP $\phi$ with $m_\phi={\cal O}(10^{-30}-10^{-29}) \, {\rm eV}$ and the initial field value $\phi_i={\cal O}(10^{17})$ GeV,
 whose late coherent oscillations
cause a tachyonic instability of the dark photon gauge fields and amplify their vacuum fluctuations to the desired strength.
Contrary to the scenario based on the conventional ALP coupling $g_{a\gamma\gamma}aF_{\mu\nu}\tilde F^{\mu\nu}$, which was explored in \cite{Majumdar:2018sbv} and \cite{Xia:2018xbt}, our scenario can be compatible with the existing observational constraints.

\begin{acknowledgments}
This work was supported by IBS under the project code, IBS-R018-D1.
We thank M. Reece for drawing our attention to \cite{Majumdar:2018sbv} and \cite{Xia:2018xbt}, and K. Kamada for useful discussions. We are particularly grateful to N. Kitajima, T. Sekiguchi, and S. Kim for helpful discussions on the lattice simulation. We also thank C. S. Shin and K. Kadota for useful comments and suggestions. 
\end{acknowledgments}


\end{document}